# Macroscopic magnetic field generated in laser atom interaction

Swarupananda Pradhan

Laser and Plasma Technology Division, Bhabha Atomic Research Centre, Mumbai-85, India

Homi Bhabha National Institute, Department of Atomic Energy, Mumbai-94, India

*Corresponding address: spradhan@barc.gov.in, pradhans75@gmail.com*

(Date: 16-08-2017)

## Abstract

We observe shift in the zero magnetic field resonance as the handedness of resonantly interacting circularly polarized light is changed. The characteristic of the shift resembles with the Zeeman light shift that arises due to interaction of non-resonant circularly polarized light with atom. However many attributes of our observed resonant phenomena like dependence on buffer gas, saturation of the shift with light intensity and involved time constant in evolution of the shift contradicts to the fictitious magnetic field model. We propose collective alignment of atomic magnetic moment giving rise to a real magnetic field as a possible mechanism behind the observed shift. The characteristic changes in the signal profile with respect to the three axis magnetic field have been established that can reveal many subtle issues pertaining to the phenomenon.

# I. Introduction

Magnetism is one of the fascinating phenomena from the beginning of contemporary science. It is ubiquitous in nature as can originate from quantum dynamics of elementary particles to giant current loop that is responsible for earth's magnetic field [1, 2]. For materials, the collective alignment of atomic magnetic moment, either spontaneously or in response to an external magnetic field is the root cause of their magnetic properties [3, 4]. So far, magnetism is extensively realized in solid and liquid state due to high atomic density. Nevertheless the gaseous phase provides a cleaner system, where internal atomic state, external degree of freedom and inter-atomic interaction can be precisely controlled; thus has opened a unique platform to address magnetism [5-10]. At the same time, coherent laser atom interaction is playing a vital role for detecting extremely weak magnetic field by pushing sensing limit to unprecedented level [11, 12].

A closely related issue is the generation of fictitious magnetic field in laser atom interaction. It gives rises to Zeeman light shift and has been treated as a part of the light shift operator

$$\delta\varepsilon = \delta\varepsilon_0 + h\delta A I \cdot J - \hat{\mu} \cdot \delta H + \delta\varepsilon_z,$$

where the terms represents centre of mass, hfs, Zeeman and tensor light shift respectively [13, 14]. The third term, Zeeman light shift affects the magnetic sub-state identical to the action of a small magnetic field $\delta H$ and exclusively arises for light with other than linear polarization. The fictitious field $\delta H$ is parallel or anti-parallel to the light propagation direction where the orientation depends on the handedness of the circular polarization as well as sign of the associated atomic magnetic moment $\mu$. A detail experimental study of the Zeeman light shift has been carried out by Cohen-Tannoudji and Dupont-Roc on Hg and Rb atoms using non resonant light generated by electrodeless discharge lamps [15]. It was shown that the Zeeman light shift vanishes at resonance, and so the assumed fictitious magnetic field. Several groups have used the assumed fictitious magnetic field to realize a variety of phenomena encompassing Sisyphus cooling [16], optical Stern-Gerlach effect [17], manipulating atomic qubits [18], interrogating cold atoms in micro traps [19] and others.

The basic theme of this paper is to investigate similar magnetic field generated in resonant laser atom interaction. The zero-field resonances observed in Hanle-type experimental procedure is an ideal tool for such studies due to inherent narrow magnetic resonances. In prior-arts, these resonances have been studied theoretically and experimentally for a variety of experimental condition. The various mechanism that contribute to the signal are optical pumping and subsequent population redistribution among Zeeman sub-state, quantum interference effect, contribution from high-rank polarization moment, transient atomic response (that can arise in a modulating magnetic field) and others [21-28]. The symmetry and polarity of the signal profile has shown dependency on details of transition, buffer gas pressure, light intensity, temperatures and others. Many instances, the problem have been treated numerically as a generalized mechanism is difficult to establish.

Here we have drawn parallel comparison between our experimental results with the predicted/verified outcome of fictitious magnetic field model. In contrary to the earlier works, our experiment is carried out at resonance where fictitious magnetic field is expected to vanish [15]. The various observations that contradict to the fictitious magnetic field model are pointed out. The distinct responses of the zero field resonance profile to the component of the magnetic field in three orthogonal directions are established and used for measurement of the generated field.

## II. Experimental method

The experiment is carried out with Rb atoms placed in 25 Torr $N_2$ buffer gas filled cell or in anti relaxation coated cell. The magnetic field at the atomic cell is controlled by using multiple magnetic shields and three set of coils in mutually orthogonal direction. A vertical cavity surface emitting laser (VCSEL) emitting 795 nm (or 780 nm) or an external cavity diode laser (ECDL) emitting 780 nm is used as the resonant light source. The schematic diagram of the experimental set-up is shown in Fig.1 [20, 21]. A part of the laser beam is used for stabilizing the laser frequency with the help of a spectrometer that represents a FM absorption spectroscopy set-up for VCSEL laser where as a saturation absorption spectroscopy set-up for ECDL. An acousto-optic modulator (AOM) is used to amplitude modulate the laser beam for a fixed on and variable off time for study pertaining to temporal evolution of the observed shift. The laser beam is either made elliptically polarized or circularly polarized by keeping the quarter-wave plate (QWP) angle at $\pm 20^0$ or $\pm 45^0$ respectively. The laser frequency is locked to one of the Rb atomic transition or at a detuned position using the spectrometer. A low frequency (55Hz) modulation is applied to the Bz magnetic field and the transmitted light through the atomic cell is phase-sensitively detected. This magnetic field modulation (MM) signal for Bz modulation is termed as MMz signal [21]. All the experimental data are taken at a residual field By=50nT, unless specified.

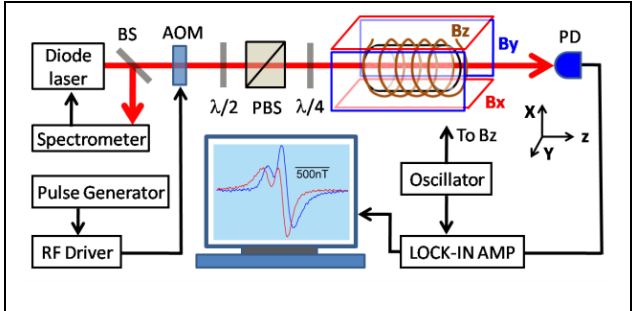

**FIG.1:** The schematic diagram with essential components for study of zero field resonance is illustrated. The diode laser represents a VCSEL laser at 795nm or 780 nm, or an ECDL at 780 nm. A small part of the laser beam is used for frequency stabilization using a spectroscopy set-up. The transmitted light passing through the atomic cell is phase sensitively detected (MMz signal) with respect to the modulation applied to the Bz field.

### III. Establishment of measurement technique

The dependence of the MMz signal on Bx and By field as a function of Bz field is shown in Fig.2 for QWP@+20$^0$. The MMz signal profile fits well with the first derivative of a Gaussian profile for Bx~ By~0 nT. It may be noted that the presented experimental profiles are extracted through modulation spectroscopy, thus represents derivative of the actual signal profile. In general, the overall profile can be approximated to

$$\frac{\partial}{\partial Bz}\left[A_1 \exp\left(-\alpha(Bz-B_1)^2\right)\right]+\frac{\partial^2}{\partial Bz^2}\left[A_2 \frac{\beta}{(Bz-B_2)^2+\beta^2}\right].$$

The second term represents the kink structure in the profile that changes its amplitude and polarity with similar changes in By field as shown in Fig.2A. As it is relatively easy to monitor small changes in By field at a residual field of By=+50nT (due to prominent kink structure), the dependence of the signal profile under this condition is studied for various Bx field (Fig.2B). In contrary to dependence on By field, the amplitude and width of the Gaussian term is found to be increasing with amplitude of the Bx field irrespective of its orientation. Thus any changes in the Bx, By and Bz field occurring due to laser atom interaction can be captured through the changes in the amplitude of the Gaussian term, changes in the kink structure and overall shift of the signal profile respectively. These distinct responses of the signal profiles are utilized for assessment of the generated magnetic field

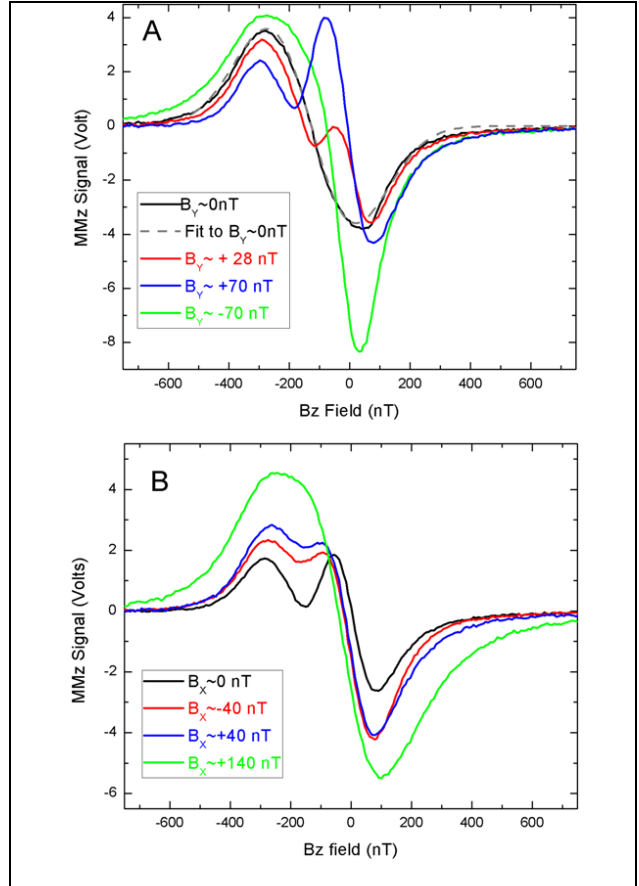

**FIG.2:** The MMz signal profiles as a function of Bz magnetic field is shown for (A) various By field with Bx=0 and (B) for various Bx field with By=50 nT. In Fig.-A, the plot for By~0nT is fitted to the first derivative of a Gaussian function (dashed gray line).

It is worth noting that even for pure circularly polarised light; the kink structure is observed for By≠0 as shown in Fig.3. Though the detail analysis is beyond the scope of this article, such a situation (where By≠0) requires transformation of the input polarization to a coordinate system along the resultant magnetic field direction [29]. Under such transformation, quantum interference can play a role even for otherwise pure circularly polarized light. However for Bx~By~0nT, the signal profile has the Gaussian term only for both QWP @ ±20$^0$ or ±45$^0$. Thus the Gaussian term has solo contribution from population redistribution at zero field level crossing, as interplay of quantum interference can be safely ruled out for pure circularly polarized light with Bx~By~0nT. A detail density matrix calculation to establish theoretical agreement to the dependence of the signal on transverse field as well as critical role of finite tilt between scanning field and laser propagation direction (not shown here) is in progress.

## IV. Results and discussions

The profile of the MMz signal shows a shift in magnetic field for reversal of handeness of elliptically and circularly polarized light interacting with Rb atoms in buffer gas filled cell as shown in Fig.3. Comparatively, the signal for linearly polarized light vanishes for same experimental parameter. However for higher amplitude of the applied magnetic field modulation, a single derivative profile (with positive slope) centred at Bz=0 is observed for linearly polarized light. We have deliberately used smaller amplitude of the modulation, as the kink structure appearing for the elliptically polarized light is not resolved at higher modulation amplitude. The shift between the signal profiles (on reversal of polarization handeness) is found to be increasing with decrease in the ellipticity of the light polarization and attain its maximum for circularly polarized light field. It resembles with the generation of the fictitious magnetic field (δH) that gives rise to Zeeman light shift in laser atom interaction. The δH reverses its orientation as the handedness of the elliptically polarized light is changed and can possibly give rise to shift in the magnetic field between the MMz signal profiles.

Though the shift in the signal has similarity with the Zeeman light shift, it may be noted that the later is an off resonant phenomena. It not only vanishes at resonance but also changes its polarity depending on the sign of the detuning [14, 15]. In contrary, the current experiment is carried out at resonance and the shift doesnot show such dependency on detuning. Another important difference is that the signal profiles for σ+ and σ- light have shown opposite polarity in the experiment related to the fictitious magnetic field [15], where as the polarity of MMz signal profiles are same for both light polarization as shown in Fig.3.

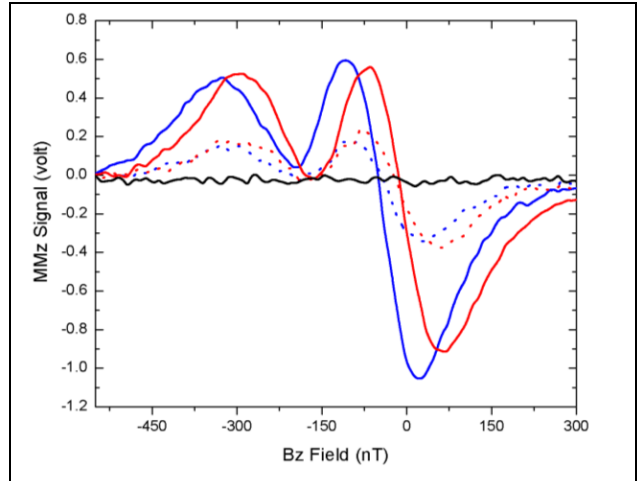

**FIG.3.** The signal profiles obtained with buffer gas filled cell are shifted for reversal of polarity of the handedness of the circularly/elliptically polarized light. The laser frequency is locked to the F=1→F'=1 transition of the $^{87}$Rb. The solid (dotted) red and blue lines corresponds to 45$^0$ and 315$^0$ (20$^0$ and 340$^0$) angle of the QWP respectively. The black line is for linearly polarized light.

The Zeeman light shift also depends on the atomic magnetic moment μ of the associated energy level. Such dependency is shown in Fig.4 for D1 transition of Rb atoms in buffer gas filled cell. For D2 transition, the contrast of the MMz signal is poor and shifts are in opposite direction (not shown here) as compared to D1 transition. These observations are well explained by the Zeeman light shift model, where the fictitious magnetic field has been shown to have opposite polarity for a ground hyperfine level coupled to D1 and D2 transitions [14, 15]. However the shifts obtained with the anti-relaxation coated cell are smaller in amplitude and have opposite polarity compared to the buffer gas filled cell as shown in Fig.4. Since the light intensity is kept same, the standard Zeeman light shift model is inadequate to explain the observation.

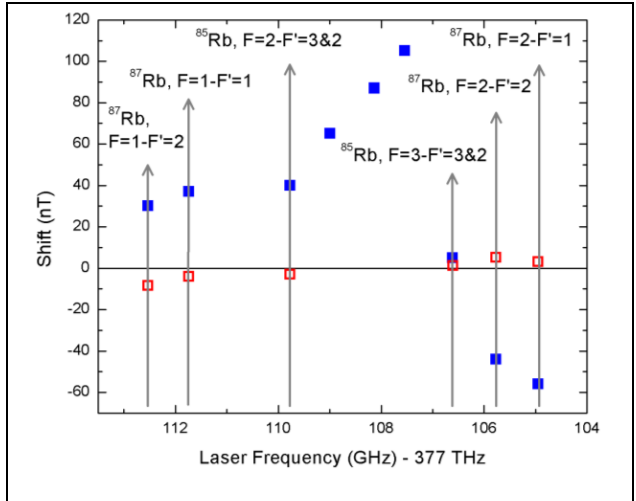

**Fig. 4.** The shifts between the signal profiles for QWP angle ±45⁰ are shown for D1 transition of Rb atoms in buffer gas filled cell (■) and in anti-relaxation coated cell (□). The polarity of the shift is reversed along with the gyromagnetic ratio. There is no data for anti-relaxation coated cell in between the Rb-85 transition due to negligible signal amplitude. The signal amplitude is highest in this region for buffer filled cell as many atoms are in resonance due to collisional broadening. Incidentally the shift is also highest in this region as shown in this figure.

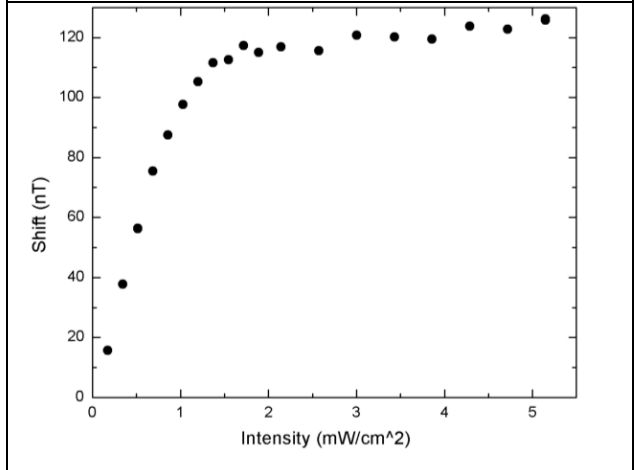

**Fig. 5.** The differential shift between the signal for ±45⁰ angles of the QWP as a function of laser intensity is shown. The laser is tuned to the D2 transition F=2→ F' of Rb-85 atoms in buffer gas filled cell. The high intensity resonant light is derived from an ECDL.

One of the important attributes of the Zeeman light shift is its linear dependence on the light intensity [15]. However, the shift observed in our experiment saturates for higher light intensity as shown in Fig.5. An ECDL locked to F=2→ F' of Rb-85 D2 transition is used as high intensity light source required for this experiment. At light intensity >5 mW/cm², the contrast of the MMz signal profile is compromised. However, the shift attains saturation well below it. The Zeeman light shift has limited explanation for the observed saturation of the shift.

In the Zeeman light shift model, the atomic energy levels are shifted due to interaction with a light field that is not linearly polarized [14]. Thus it is an instantaneous process with respect to application of the light field. Fig.6 shows variation in the shift on the parameters of an amplitude modulated light beam. The experiment is carried out with an amplitude modulating VCSEL laser beam generated by use of an AOM. The Ton period is fixed at ~1, 5, 10 or 20 µs, whereas the Toff period is varied from 0 to 100 µs. The first order diffracted beam from the AOM is used to ensure complete switching off of the light beam during Toff period. The interpretation of data involves intricacy of evolution of the shift during Ton and decay during Toff period in multiple pulses. However, for larger Toff it will converge to the average evolution during single pulse. Nevertheless an instantaneous process would provide a fixed shift irrespective of the Ton or Toff period. Conclusively, the observed shift is not an instantaneous process and a finite time constant is involved in its evolution.

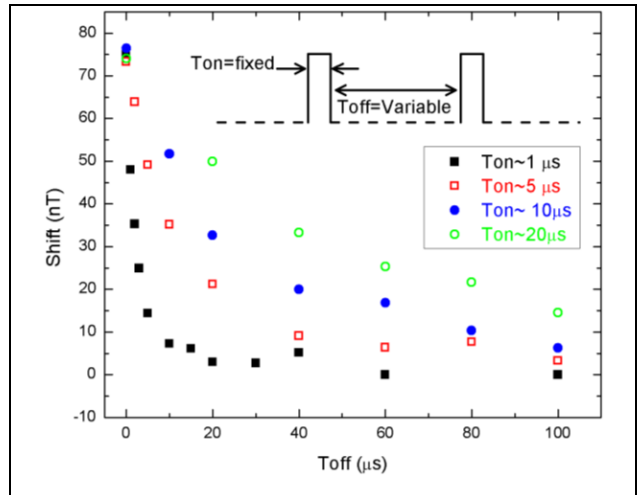

**Fig. 6.** Dependence of the shift on the switching off time (Toff) of the laser beam for a fixed switched on time (Ton) ~1, 5, 10 and 20 µs are shown. As the Ton is increased, the shift is preserved for a longer Toff time. The laser is locked to the F=2→ F' transition of ⁸⁵Rb.

The observed shift at resonance, increase of the shift with decrease in ellipticity, enhanced shift for buffer gas filled cell compared to anti-reflection coated cell, saturation of the shift at higher intensity, associated time constant involved in the evolution of the shift can be explained by assumption of a real magnetic field that is produced by alignment of the atomic magnetic moment $\mu$ during laser-atom interaction. Such kind of collective alignment is realized in ferromagnetic solid state material due to exchange interaction. The quantum mechanical exchange interaction decays exponentially and competition with the relatively long range ($1/r^3$) dipolar interaction is compromised by the domain formation in ferromagnetic materials [3, 4]. Since the gaseous sample is exceedingly dilute, the actions of these forces are extremely feeble. At the same time the gaseous sample is void of detrimental phonon waves. Though the direct exchange interaction may have limited role in dilute gas phase, the high density buffer gas can play critical role through *super-exchange* interaction to facilitate collective alignment [3]. The buffer gas can also influence by confining the atoms for longer duration in the interaction volume, thereby providing requisite time for alignment of the atomic magnetic moments. Thus buffer gas filled cell can have advantages for possible collective alignment of atomic magnetic moment over anti-relaxation cell.

Similar to the assumption made for fictitious magnetic field, it is assumed that the atomic magnetic moments are aligned parallel to the light field to produce a real magnetic field. The orientation of the generated magnetic field is changed as the polarity of the atomic magnetic moments is changed. For circularly polarized light, the atomic population is optically pumped to the extreme ground state ($N_e$). The generated magnetic field and hence shift (S) in the resonance profiles will be $S \propto N_e - N/(2F+1)$, where $N$ and $F$ are total atomic population and ground hyperfine quantum number respectively. A simple rate equation for the population $N_e$ (while incorporating steady state condition) can be written as

$$\frac{dN_e}{dt} = \alpha(N - N_e)I - 2F\beta N_e + \beta(N - N_e)$$

Where $\alpha$ and $\beta$ are related to optical pumping rate and population redistribution rate respectively, and $I$ is laser intensity. Using initial condition at t=0, the equation will lead to a solution

$$N_e(t) = \frac{N}{\alpha I + (2F+1)\beta}\left[\alpha I + \beta - \frac{2F\alpha I}{2F+1}\exp[-[\alpha I + (2F+1)\beta]t]\right].$$

The parametric dependences of Ne/N (that is proportional to the shift) are shown in Fig.7 for $\alpha = \beta = F = 1$. The illustrated plots suitably explain the observations made in Fig. 3, 5 and 6. In connection with Fig.7C, the $\alpha$ can be practically changed by changing the ellipticity of the light beam. This explains the increase in the shift as the light polarization changed from elliptical to circular (Fig.3).

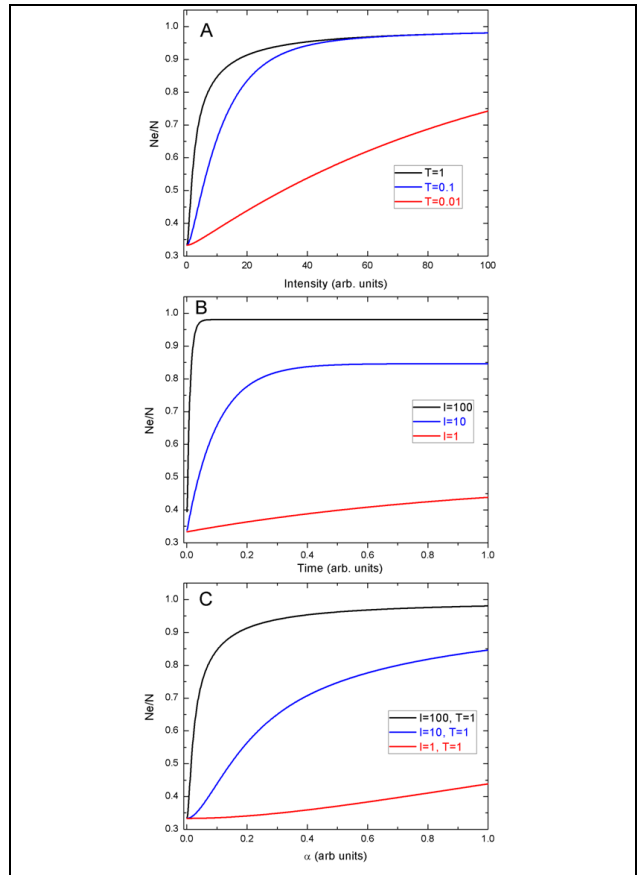

**Fig.7**. The fraction of atoms in the extreme state as a function of laser intensity, time and optical pumping rate are calculated from the solution of the rate equation. (A) The saturation with the laser intensity explains Fig.5. (B) The associated time constant in evolution of the shift is consistent with the experimental results shown in Fig. 6. (C) The increase in the shift with decrease in the ellipticity as the polarization changed from elliptical to circularly polarized light in Fig.3 is consistent with the shown $\alpha$ dependence.

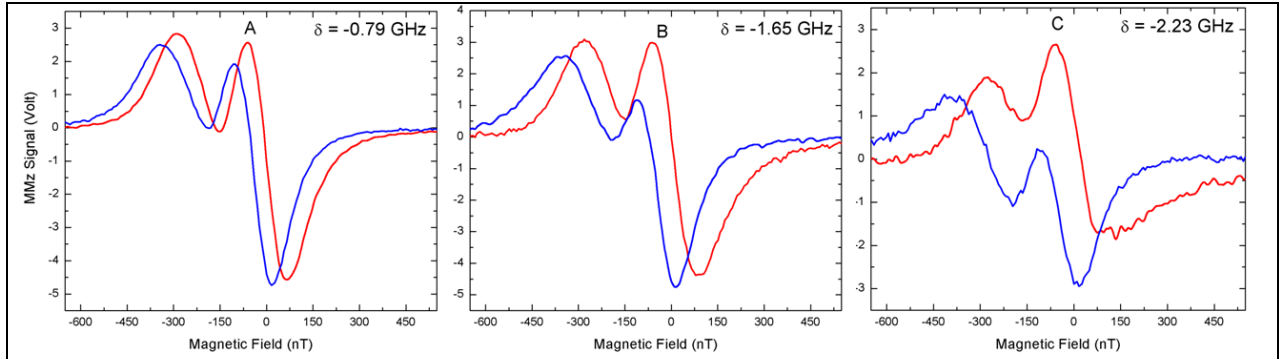

**FIG. 8.** The laser is locked at different detuning from F=2→F'=2, 3 transition of $^{85}$Rb atom. The red (blue) experimental curve is for QWP angle of $20^0$ ($340^0$). The shifts among the profiles and the change in kink amplitude indicate variation in mean field along z and y-direction respectively. The shifts for these plots are shown in Fig.4.

The proposed model involving generation of a real magnetic field has solely explained many of the observed phenomena. However, for laser locked in between the $^{85}$Rb F=2→F'=2, 3 and F=3→F'=2, 3 transition, the measured shifts are larger than the values at resonances (Fig.4). The corresponding signal profiles are shown in Fig. 8. In this laser frequency regime, different group of atoms in ground levels F=2 as well as 3 are off resonantly coupled with the laser field due to large homogeneous broadening caused by the high buffer gas pressure. Since the gyromagnetic ratios of the involved ground levels have opposite polarity, a smaller shift between the spectral profiles is expected. There is also a possible interplay of fictitious magnetic field in this case due to coupling of the off-resonant light field. As the light field is detuned in opposite direction and also the gyromagnetic ratio are opposite, the fictitious magnetic field for both the species will be in the same direction leading to larger shift. However the temporal evolution of the shift in this regime is found to be associated with a time constant similar to Fig.6, thus rules out the interplay of fictitious magnetic field. Further, the cancellation of shift do occur at F=3→F' transition of the $^{85}$Rb atoms. The basis for cancelation of the shift at the frequency position corresponding to F=3→F' transition of the $^{85}$Rb atoms rather than in between the transition of D1 lines is in progress. One of the possible reasons may be due to dipolar interaction between the two different species that are produced in this frequency regime. In fact, such interaction will produce a torque on each species leading to generation of a magnetic field in perpendicular direction. This might be the reason for the large change in amplitude of the kink structure as shown in the Fig.8.

## V. Conclusions

A three axis magnetic field measurement technique is established to monitor magnetic field generated during laser atom interaction. The attributes of the observed shift is compared with the Zeeman light shift arising due to fictitious magnetic field. The limitations of the Zeeman light shift in explaining the phenomena are pointed out. The involved time constant in the evolution of the shift and other observations contradicts with the fictitious magnetic field model. A model involving generation of a real magnetic field that suitably explain the observation is proposed. The rate equation based on the assumption of generation of a real magnetic field (due to collective alignment of atomic magnetic moment) has explained the observed behaviour.

**Acknowledgments:**

We are extremely thankful to R. K. Rajawat, Asso. Director, BTDG for supporting this program.